\documentclass[iop,amssymb,amsfonts,amsmath]{emulateapj}

\usepackage[bookmarks,bookmarksopen,colorlinks,linkcolor={red},citecolor={red},urlcolor={blue}]{hyperref}

\usepackage{graphicx}
\usepackage{bm}
\usepackage{hyperref}
\usepackage{color}
\usepackage{subfigure}

\def\gtsima{$\; \buildrel > \over \sim \;$}
\def\ltsima{$\; \buildrel < \over \sim \;$}
\def\gsim{\lower.5ex\hbox{\gtsima}}
\def\lsim{\lower.5ex\hbox{\ltsima}}
\def\simleq{\; \raise0.3ex\hbox{$<$\kern-0.75em \raise-1.1ex\hbox{$\sim$}}\; }
\def\simgeq{\; \raise0.3ex\hbox{$>$\kern-0.75em \raise-1.1ex\hbox{$\sim$}}\; }

\def\apjl{ApJL}
\def\apjs{ApJS}
\def\aap{A\&A}

\newcommand{\alf}{Alfv\'en}

\begin{document}

\title{Cosmic Ray propagation in Galactic turbulence}
\author{Carmelo Evoli\altaffilmark{1} and Huirong Yan\altaffilmark{2}}

\altaffiltext{1}{{II.}~Institut f\"ur Theoretische Physik, Universit\"{a}t Hamburg, Luruper Chaussee 149, D-22761 Hamburg, Germany; carmelo.evoli@desy.de}
\altaffiltext{2}{Kavli Institute of Astronomy and Astrophysics, Peking University, Beijing 100871, China; hryan@pku.edu.cn}


\begin{abstract}
We revisit propagation of galactic cosmic rays (CRs) in light of recent advances in CR diffusion theory in realistic interstellar turbulence. 
We use a tested model of turbulence in which it has been shown that fast modes dominate scattering of CRs. As a result, propagation becomes inhomogeneous and environment dependent. 
By adopting the formalism of the nonlinear theory developed by Yan \& Lazarian, we calculate the diffusion of CRs self-consistently from first principles. 
We assume a two-phase model for the Galaxy to account for different damping mechanisms of the fast modes, and we find that the energy dependence of the diffusion coefficient is mainly affected by medium properties.  
We show that it gives a correct framework to interpret some of the recent CR puzzles.
\end{abstract}


\maketitle 

\section{Introduction} 
Understanding galactic cosmic ray (CR) propagation is a crucial topic in astrophysics. 
CRs are a unique probe of the interstellar medium (ISM) properties since they can transverse extended regions in the Galaxy before reaching the Earth's atmosphere providing information about galactic magnetic fields, gas distributions and stellar rates. 
Moreover, new and upcoming detectors (e.g., AMS-02 \footnote{\url{http://www.ams02.org}}, CALET \footnote{\url{http://calet.phys.lsu.edu/}}, ISS-CREAM \footnote{\url{http://cosmicray.umd.edu/iss-cream-home.html}}) are expected to provide CR spectral data with unprecedented precision which will require an accurate description of the CR interactions with the ISM. 
Finally, CRs could be the first place where the elusive Dark Matter component of the universe will be detected~\citep{Bertone:2010at,Bergstrom:2012fi}.

According to the standard scenario, the bulk of the observed spectra of CRs are shaped by two basic processes:
the acceleration in the galactic supernova shocks and the following propagation in the ISM~\citep{Ptuskin:2012,Yan:2012}. 
In fact, after being accelerated in the sources, the charged energetic particles of the cosmic radiation diffuse in the turbulent galactic magnetic field  that is responsible for their high isotropy and longer confinement time in the Galaxy with respect to ballistic crossing time. 

The propagation of galactic CRs is usually described in terms of a diffusion equation~\citep{Berezinskii:1990}:
\begin{equation}\label{Eq:diffusion}
\frac{\partial N(\vec{r},p,t)}{\partial t} - \nabla (D_{xx} \nabla N) = Q(\vec{r},p,t)
\end{equation}
where $N(\vec{r}, p, t)$ is the time-dependent CR density per unit of total particle momentum $p$ at the galactic position $\vec{r}$ and $Q(\vec{r},p,t)$ incorporates energy loss processes in the ISM, nuclear fragmentation, radioactive decay of unstable nuclei, and the properties of CR sources. 
$D_{xx}$ is the spatial diffusion tensor and, in the more general case, is anisotropic and a function of position. 
On the microscopic level, the spatial diffusion of CRs results from the resonant and nonresonant (transit time damping, or TTD) interaction of CRs with galactic MHD turbulence~\citep{Schlickeiser:2002}.
The solution for the diffusion equation is usually obtained under steady-state assumption (i.e.,~$\partial N/\partial t \rightarrow 0$), since source properties are assumed to be constant during the diffusion time scale of $\sim$GeV CRs ($\sim 10^8$~yr as inferred from unstable secondary nuclei observations).  

At present, propagation of CRs is an advanced theory which makes use of both analytical studies and numerical simulations.  
Thanks to these joint efforts, substantial progress has been made in understanding MHD turbulence during past decades. 
According to the current scenario, MHD turbulence is composed of anisotropic \alf~modes ($k_\bot \gg k_\|$, \citealt{GS95}, henceforth GS95)
\footnote{In the present work, the directions perpendicular ($\perp$) and parallel ($\|$) are always referred to with respect to the magnetic field.}
as well as isotropic fast modes as both theoretically demonstrated and numerically confirmed by different simulations~\citep[see, e.g,][]{Cho:2002qi,Kowal:2010iw} (see also \citealt{CLV_lecnotes} for a review). In addition, the observations from solar wind also support the GS95 picture of \alf ic turbulence \citep[see, e.g.,][]{solarwind_turb}. 

This progress inevitably leads to the corresponding paradigm shift in the CR propagation theory, which is closely linked to the models of turbulence.
Scattering efficiency is many orders of magnitude lower with the \alf~modes than earlier predictions made with ad hoc models because its scale dependent anisotropy and fast modes were identified as the dominant scattering agent for CRs~(\citealt{Yan:2002,Yan:2004}).

\begin{figure}[thbp]
\centering
\includegraphics[width=0.48\textwidth]{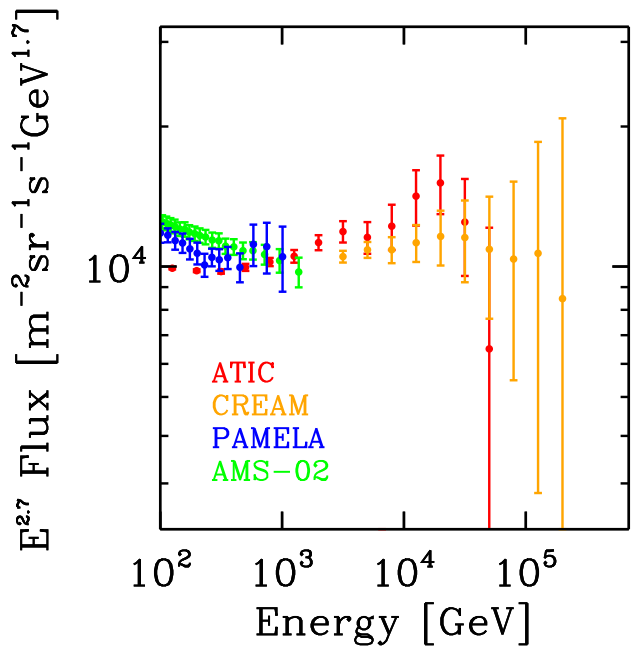}
\caption{Proton absolute fluxes measured above 100 GeV. See Table~\ref{Tab:data} for a reference list of the experimental data.\\}
\label{Fig:protons}
\end{figure}

Fast modes are not only much more efficient in gyroresonance interaction but also dominate the scattering for most of the pitch angle range (including $90^\circ$) through the TTD interaction according to nonlinear theory (NLT)~\citep{Yan:2008}, which is confirmed by test particle simulations~\citep{Xu:2013fk}. 
Finally, scattering by fast modes naturally results in inhomogeneous diffusion, since it is determined by medium properties as first predicted by~\citet{Yan:2002,Yan:2004}. 

In spite of the impressive theoretical work done in this direction, numerical and semi-analytical models, developed for solving the CR diffusion equation in the more realistic conditions of ISM, are all based on the earliest turbulence models. 
In fact, the diffusion coefficient in the empirical slab turbulence with spectrum $W(k) \propto 1/k^{2-\delta}$ for $k>k_L$ is given by $D \propto v \rho^\delta$, where $\rho = p/Z$ is the rigidity and $v$ is the velocity. 
The usual formalism for wave--particle interactions used to derive this result is quasilinear theory (QLT), in which the turbulent magnetic field is assumed to be negligible with respect to the regular (on a galactic scale) component.

The value of $\delta$ can be deduced from the observed secondary to primary, e.g., boron/carbon (B/C), ratio in the high-energy CR fluxes. 
The total column density of matter ($X = n_{0}\tau v$) they penetrate during their residence time, $\tau$, in a medium with target number density $n_0$ can be expressed in terms of the system size, $L$ (usually it is assumed the thickness of the galactic halo), and the spatial diffusion coefficient:
\begin{equation}
X(\rho) = \frac{3n_{0}L^{2}}{\lambda_{||}(\rho)} = \frac{n_{0}vL^{2}}{D(\rho)} \, ,
\end{equation} 
where the mean free path is defined as $\lambda_{||} = 3D/v$. 
Combining it with the equation for $D=D(\rho)$, it is easy to draw the conclusion that $X \sim \rho^{-\delta}$.
\begin{figure}[thbp]
\centering
\includegraphics[width=0.48\textwidth]{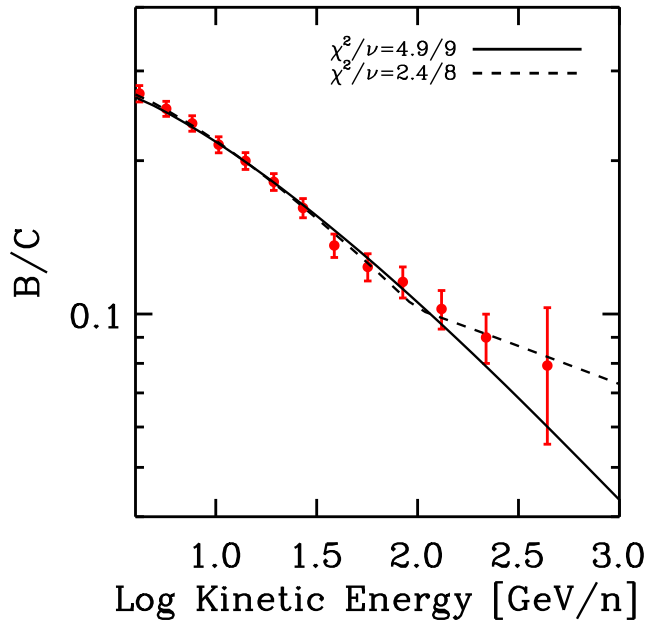}
\caption{B/C ratio from AMS-02 compared to our best-fit models: single power-law (solid), broken power-law (dashed). The best-fit reduced chi-square against AMS-02 data is reported.\\}
\label{Fig:break}
\end{figure}
The observed decrease of the B/C ratio at energies above 1~GeV~n$^{-1}$ constrains $\delta$ to be in a range of around $0.5-0.6$, which can not be reconciled with the Kolmogorov power spectrum $\delta=1/3$ as is the case for the ISM, as inferred by radio scintillation and refraction observations~\citep{Lee1976,Armstrong:1995}.

A second problem in modeling galactic propagation is the role of re-acceleration. 
Weak distributed stochastic re-acceleration by interstellar MHD turbulence 
seemed to be a natural effect to be implemented in propagation models in order to obtain a better fit of the B/C peak at $\sim 1$~GeV~n$^{-1}$~\citep{Seo:1994}. 
However, in order to be significant, re-acceleration requires an interstellar \alf~velocity of $\sim 10-30$~km s$^{-1}$ that, for a density of the ionized ISM component of $\sim 1$~cm$^{-3}$~\citep{Ferriere:2001rg}, corresponds to an
average magnetic field of $~10-30\, \mu$G over the CR propagation region, whereas magnetic fields in the solar neighborhood are observed to be $\sim 2$~$\mu$G for the regular field and $\sim 3$~$\mu$G for the random field~\citep{Beck:2013}.
Moreover, recent analysis of the diffuse synchrotron emission by CR lepton in the galactic magnetic field allowed the probe of the {\em interstellar} lepton spectrum which has been found to be incompatible with the features expected in re-acceleration scenarios~\citep{Jaffe:2011,DiBernardo:2013}.

Both of the mentioned difficulties are a possible consequence of assuming a diffusion coefficient as a single power-law in rigidity based on the ad hoc description of MHD turbulence. 
In addition, estimates of the interstellar magnetic field fluctuations~\citep{Jansson:2012rt,Jansson:2012pc}
from Faraday rotation measurements and radio polarization studies indicate rather comparable levels of the turbulent and regular components, 
which invalidate the QLT assumption adopted, particularly for treating the nonresonant interaction, TTD, between galactic CRs and turbulence.
In view of the challenge that recent observations posed to conventional homogeneous CR diffusion theory, it is of utter importance to incorporate the revised diffusion theory for tested models of turbulence to the modeling of CR propagation. 

\begin{table*}[htdp]
\caption{References for the Experimental Data Used in This Work.}
\begin{center}
\begin{tabular}{ccccc}
\hline
Name of the Experiment & Type & Data & Years of Data Taking & Reference \\
\hline
ATIC         & Balloon & Proton flux & 2002-2003 & \citealt{ATIC2} \\ 
CREAM-II & Balloon & Proton flux & 2005-2006           & \citealt{CREAM} \\
PAMELA   & Satellite & Proton flux & 2006-2008 & \citealt{Adriani:2011cu} \\
AMS-02    & Satellite & Proton flux & 2013 & ICRC 2013 contribution: 1265 \\
HEAO-3   & Satellite & B/C   & 1979-1980     & \citealt{HEAO3} \\
CREAM-I  & Balloon & B/C    & 2004-2005 & \citealt{Ahn2008} \\ 
CRN         & Satellite & B/C  & 1985     & \citealt{CRN} \\
PAMELA   & Satellite & B/C  & 2006-2008 & ICRC 2013 contribution: 0538\\
AMS-02    & Satellite & B/C & 2013  & ICRC 2013 contribution: 1266\\
\hline
\end{tabular}
\end{center}
\label{Tab:data}
\end{table*}%

\section{A break in diffusion?}
Galactic CRs are assumed to be accelerated in astrophysical sources, such as supernova remnants (SNRs), with a source spectrum of $Q_{\rm CR} \propto E^{-\gamma}$.
In fact, at energies larger than $m_p$, the power law behavior (with $\gamma \sim 2-2.2$ in the case of strong remnant shocks), naturally arises from diffusive shock acceleration theory~\citep{Malkov2001,Caprioli2008}.
It is straightforward to see from Equation~\ref{Eq:diffusion} (neglecting energy losses and nuclear interactions) that the nuclei spectra observed from Earth after propagation
have to be a single power-law, in particular, $\propto E^{-\gamma-\delta}$, for energies $\gg 1$~GeV~n$^{-1}$.
This result is at odds with recent measurements by the PAMELA experiment that showed a change of slope at $\sim 230$~GV, e.g., for protons, 
from $\propto E^{-2.85}$ for $E < 230$~GeV to $\sim E^{-2.67}$ for $E > 230$ GeV~\citep{Adriani:2011cu} for the proton and helium spectra. 
A change in the proton slope at high-energy is also consistent with high-accuracy balloon measurements (ATIC-2 and CREAM) at energies from $\sim 10$ to $\sim 10^5$~GeV. 

More recently, the AMS-02 collaboration reported accurate measurements of the proton flux up to 1.8~TeV.
In the high energy region above 100 GeV probed by this experiment the spectrum is consistent with a single power-law spectrum and shows no fine structure or break, leading to the conclusion that the hardening required to reconcile ATIC and CREAM data must be at higher energies (see Figure~\ref{Fig:protons}).

Among the proposed explanations, a high-energy break in primary CR fluxes can be easily reproduced with a change in the diffusion coefficient single power-law behavior. A more natural observable that can be used to confirm such a scenario is any secondary over primary ratio, for example, the anti-proton over proton ratio or B/C. As shown in~\citet{Evoli:2011id}, those ratios are independent of source properties and depend almost only on diffusion properties at high-energies. 

However, before AMS-02 data, it was not possible to perform this analysis, since the data available at that time lay in a range of lower energies.

In order to investigate the presence of a break in diffusion from the B/C data, we first assume that the diffusion coefficient can be approximated as a single power-law $D \propto E^{\delta}$ for energies $> 5$~GeV n$^{-1}$. A second possibility is that the diffusion coefficient changes its slope above a specific rigidity ($230$~GV) where it becomes $D \propto E^{\delta_{\rm H}}$.

In Figure~\ref{Fig:break}, we show the best fit obtained for the two different cases by solving the diffusion equation in~\ref{Eq:diffusion} for a minimum energy of $\gsim 5$~GeV (a motivation for this minimum energy can be found in \citealt{Evoli:2008dv}).
When a single power-law for the diffusion equation is assumed, a value of $\delta=0.44$ is obtained, while in presence of the break, the slope changes from $\delta=0.46$ to $\delta_{\rm H}=0.17$.
The second case is fitted with a slightly better reduced chi-square, 
however, not statistically relevant to confirm the presence of a break in the diffusion coefficient.

In the following section, we show how our model can easily account for a high-energy break in diffusion, even if more data are required to clarify the emerging picture.

\section{NLT diffusion in tested model of turbulence}
\begin{figure*}[tbp]
\centering
\includegraphics[width=0.46\textwidth]{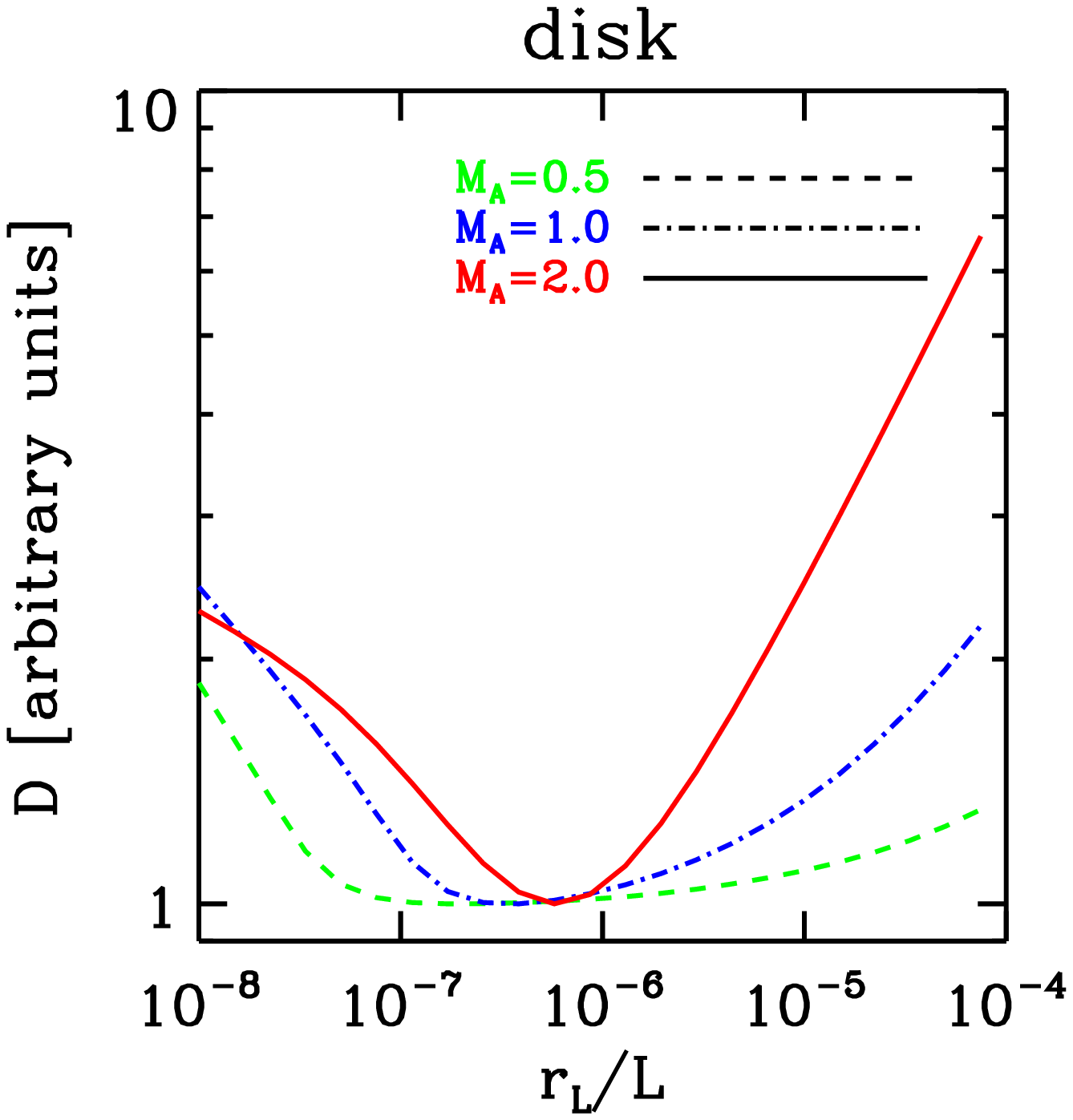} \, \, \, \, 
\includegraphics[width=0.46\textwidth]{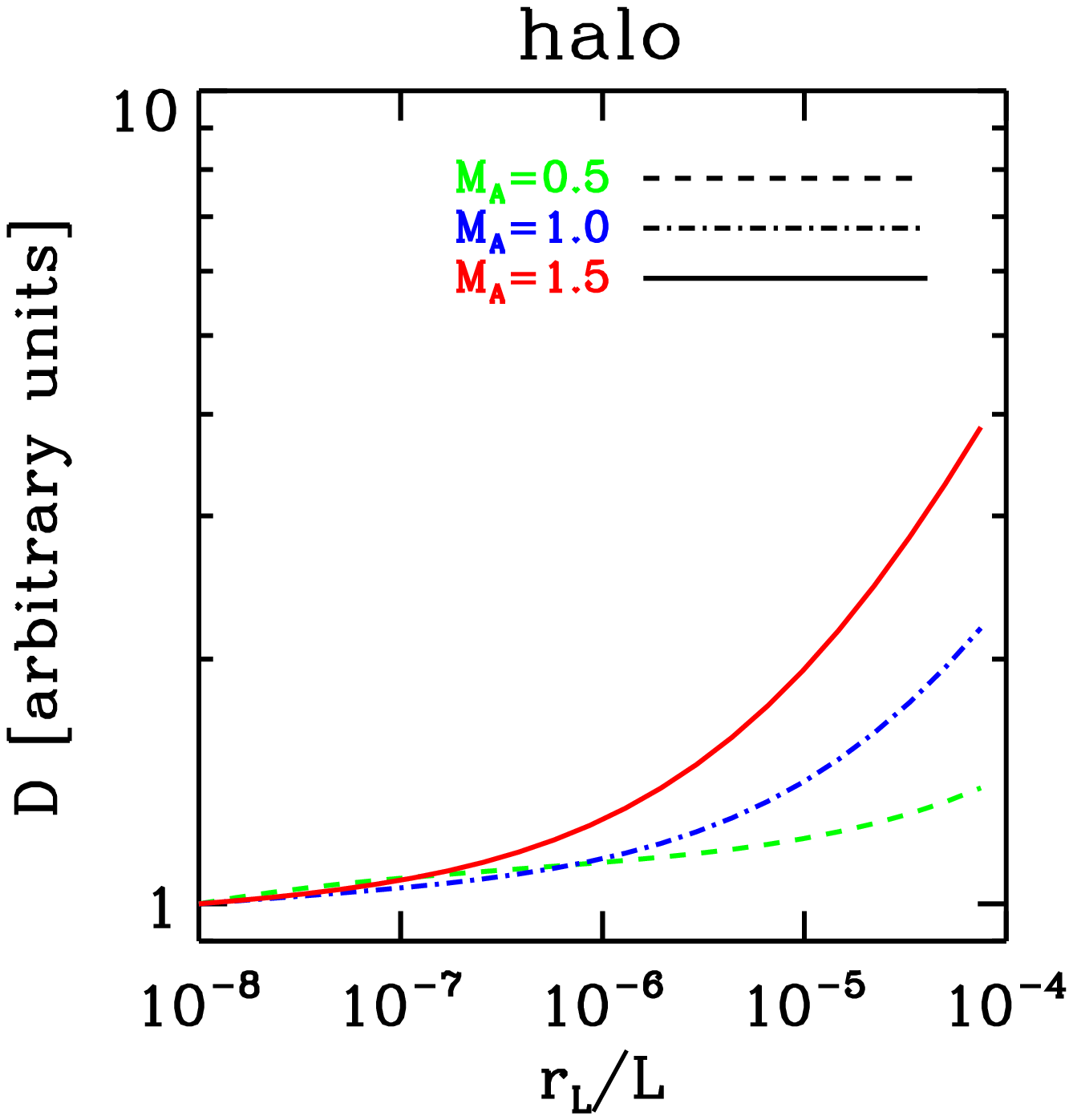}
\caption{Diffusion coefficient as a function of the Larmor radius in different phases of the ISM: disk (left) and halo (right) for different values of $M_A$. Particle rigidity is obtained by multiplying $r_L$ to the local magnetic field intensity.}
\label{Fig:D}
\end{figure*}
A self-consistent picture of galactic CRs propagation can be achieved on the basis of a numerically tested theory with solid theoretical foundations.
Interstellar turbulence is usually considered to be injected at spatial scales of the order of $\sim 10^{20}-10^{21}$~cm, as a result of supernova explosions. The following turbulence cascade transfers the turbulent energy to smaller spatial scales through cascade. On small scales, the compressible MHD turbulence can be decomposed into \alf ic, slow, and fast magnetosonic modes~\citep{Cho:2002qi}. 
Among them, the GS95 scaling applies to the \alf ic and slow magnetosonic modes~\citep{Lithwick:2001,Cho:2002}. 
In that case, the turbulent energy is preferentially cascaded in the direction perpendicular to the magnetic field, and this leads to strong suppression of relativistic particle scattering.
Conversely, the cascade of fast magnetosonic modes is isotropic with the Iroshniokov-Kraichnan ($I_{F} \propto k^{-7/2}$) scaling~\citep{Cho:2002,Cho:2003dd}, and fast modes were shown to have the dominant contribution to the scattering of CRs in the ISM~\citep{Yan:2002,Yan:2004,Yan:2008}.

To calculate NLT diffusion in the different environments of the ISM and to address the problem of perpendicular transport, we refer to the results obtained in~\cite{Yan:2008}.
We recap here the main assumptions and their results.

In contrast to QLT in which an unperturbed orbit of the scattered particles is assumed, NLT accounts for the gradual variation of the particle pitch angle ($\mu$) with the magnetic field ($B$) in compressible turbulence due to the first adiabatic invariant, leading to a Gaussian broadening of the resonance function: 
\begin{equation}
R_{n}^{\rm NLT} ( k_{\parallel} v_{\parallel} - \omega \pm n \Omega ) 
= \frac{\sqrt{\pi}}{k_{\parallel}\Delta v_{\|}} \, {\rm exp} \left[ {-\frac{(k_{\parallel}v \mu-\omega \pm n\Omega)^{2}}{k_{\parallel}^{2} \Delta v_{\|}^{2}}} \right]
\end{equation}
where $\Omega$ and $\omega$ are the Larmor frequency and the wave frequency of the CRs, respectively, $\Delta v_{\|}$ is the average uncertainty of the particle parallel speed caused by the magnetic perturbations
$\tilde{B}_{\|}$ and can be approximated as $\Delta v_{\|} / v_{\perp} \sim \langle \tilde{B}_{\|}^{2} \rangle / B_{0}^{2}$.

The corresponding pitch angle diffusion can then be calculated from:

\begin{equation}\label{Eq:Dmm}
D_{\mu\mu} = \frac{\Omega^{2} (1-\mu^{2})}{B_{0}^{2}} \int \!\!\! d^{3}k \, R^{\rm NLT}_{n}({\mathbf k}) \left[\frac{k_{\parallel}^{2}}{k^{2}} J_{n}^{\prime 2}(w) I^{F}({\mathbf k})\right]
\end{equation}

where $w \equiv k_{\perp} v_{\perp} / \Omega$ and $J_{n}$ represents the Bessel function and we neglect the contribution from \alf ic modes because of their anisotropy as discussed above.

Unlike \alf ic turbulence, magnetosonic modes are subjected to various damping processes that could halt the cascade. 
Scattering by fast modes is, therefore, influenced by the medium properties, which determines the damping.
We consider here two different regions in the Galaxy: 
the {\it halo} in which collisionless damping is dominant and the {\it disk} in which viscous damping is additionaly taken into account.
The cutoff scale $k_c$ due to damping can be obtained by equating the cascading rate of fast modes with the relevant damping rate.
In the case of collisionless damping:
\begin{equation}
k_{c}L = \frac{4M_{A}^{4}\gamma\xi^{2}}{\pi \beta (1-\xi^2)^{2}} {\rm exp} \left( \frac{2}{\beta \gamma \xi^{2}} \right)
\end{equation}
where $M_{A} \sim \tilde{B} / B$ is the \alf ic Mach number, $\gamma \equiv m_{p}/m_{e}$ is the ratio between proton and electron mass, and $\beta \equiv P_{\rm gas}/P_{\rm mag}$ is the ratio between thermal and magnetic pressure in the ISM. 
Note that the scale $k_{c}$ depends on the wave pitch angle $\xi$, which makes the damping anisotropic. 
In the disk the Coulomb collisional mean free path is $l_{\rm mfp} \sim 6 \times 10^{12}$~cm and $\beta \sim 0.1$, and the viscous damping cut-off scale can be evaluated as:
\begin{equation}
k_{c}L = x_{c}(1-\xi^{2})^{-2/3}
\end{equation}
%
%
where
%
$x_{c} \equiv \left( 6 \rho R_{m}/ v_{A} \right)^{2/3}$ is a combination of the following parameters: the \alf~velocity $v_{A}$, the magnetic Reynolds number $R_{m}$, and the medium density $\rho$.
For values of these parameters typical of the warm ionized component of the ISM, e.g., in~\citet{Ferriere:2001rg}, $x_{c}$ is of the order of $10^{6}$.   

Equation~\ref{Eq:Dmm} can be specified for gyro-resonance ($D_{\mu\mu}^{\rm G}$, corresponding to $n\ne 0$) and TTD ($D_{\mu\mu}^{\rm T}$ for $n=0$).
TTD arises from Landau type interactions of particles with the compressive component of magnetic fluctuations (i.e., the component parallel to the mean magnetic field $B_{0}$).

Finally, we can compute the spatial diffusion coefficient by means of the following expression:
\begin{equation}
D \sim \frac{1}{3} \lambda_{||} v = \frac{1}{8} \int_{-1}^{1} d\mu \frac{ v (1-\mu^{2})^{2} }{ D_{\mu\mu}^{\rm G}+D_{\mu\mu}^{\rm T} }  
\end{equation}

In Figure~\ref{Fig:D} we show the diffusion coefficient as a function of the particle rigidity ($r_{L}$ is the particle Larmor radius) for different values of the level of turbulence expressed by $M_{A}$. 
In the disk-like environment, for a very turbulent medium $M_{A} > 1$, the diffusion coefficient exhibits different behaviors above and below the critical rigidity, $r_L / L \sim 10^{-6}$, which corresponds\footnote{In the relativistic limit: $r_L \sim \frac{A}{Z}\left( \frac{E}{10^{15}{\rm eV}} \right) \left( \frac{B}{1 \mu G}\right)^{-1} {\rm pc}$} to a kinetic energy per nucleon of $\sim 1$~GeV assuming $B\sim 1\mu$G and $L=10$~pc) and a dependence $D \sim E^{0.5}$ above the break, as required to explain the observed high-energy B/C ratio. 
The observed energy dependence is mainly due to the different behavior with energy of the damping scales as first proposed in \citet{Yan:2002}. 
Diffusion in the halo is a monotonic increasing function of the energy, given by the fact that collisionless damping is always dominant. Depending on the turbulence level the diffusion coefficient can be approximated as $\sim E^{0.3-0.4}$ at higher energies. 

In general, larger magnetic turbulence corresponds to more efficient diffusion through the collisionless damping scale.
In~\cite{Evoli:2012}, a similar trend has been proposed to account for the mismatching between the inferred CR source distribution from the galaxy diffuse maps and the SNR observed distributions.

We implement the diffusion coefficients derived for the different galactic environments in the~\href{http://www.dragonproject.org}{DRAGON} code~\citep{Evoli:2008dv} to evaluate CR propagation on a galactic scale.
The numerical code solves equation~\ref{Eq:diffusion} in the steady-state limit defined as $\partial N / \partial t \rightarrow 0$, taking into account an accurate description of CR source, the relevant energy-losses for nuclei in the ISM, and gas density distributions. 
In particular, the latter is relevant for secondary production. 
DRAGON assumes a diffusion zone with cylindrical symmetry, within which CRs diffuse and beyond which they escape, with a radius of $R=20$~kpc and scale height of $L=4$~kpc.
We assume CRs propagate in the disk unless their mean displacement ($\sim \sqrt{D\tau}$) exceeds the scale height of the warm ionized ISM component ($h\sim 1$~kpc; \citealt{Cordes:1991}), in which case we assume halo diffusion. This condition defines the critical rigidity at which the transition between propagation in the disk and in the halo takes place. 
Given that the typical value of the diffusion coefficient found from the fit to CR data is $D\sim 3\times 10^{28}$~cm$^{2}$~s$^{-1}$ at energy $\sim1$~GeV nucleon$^{-1}$ and $\tau \sim 10^7$~yr, it is easy to derive that the critical rigidity has to be $O(100)$~GV.

\begin{figure}[t]
\centering
\includegraphics[width=0.48\textwidth]{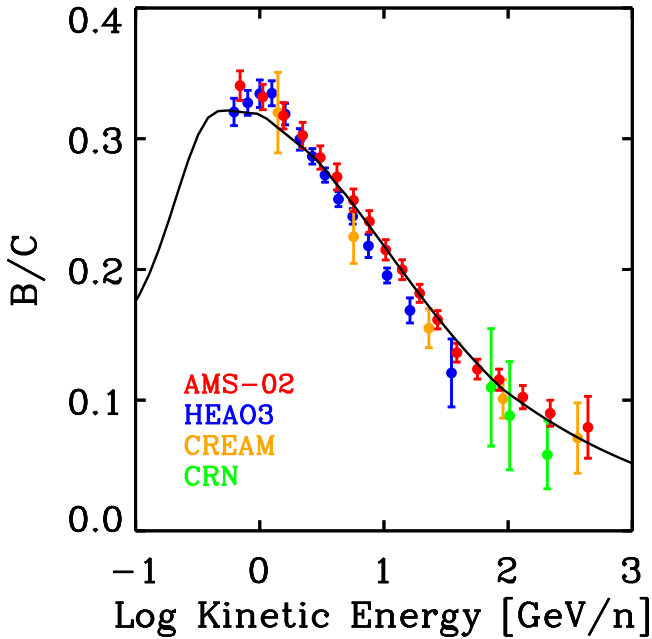}
\caption{Comparison of our model with $M_A = 2$ for the disk and $M_A = 1$ in the halo and modulated with a 100 GV potential against B/C data. See Table~\ref{Tab:data} for a reference list of the experimental data.\\}
\label{fig:data}
\end{figure}

In Figure~\ref{fig:data}, CR spectra obtained in our model, assuming $M_{A,\rm disk}=2$ and $M_{A,\rm halo}=1$, are plotted against the B/C ratio. In order to account for solar modulation, we assume the force-field approximation with a small potential (100 MV) that can safely reproduce a more realistic charge-dependent model, as the one presented in~\citet{Maccione:2012cu}.   
Noticeably, the $\sim1$~GeV break in the B/C can be reproduced without introducing re-acceleration in the propagation model.
According to our scenario, the spectral break observed in high-energy CR data is an effect of the change in the turbulence properties of the ISM as seen by CRs of different energies.

A different explanation for the break has been recently proposed by~\citet{Blasi:2012yr} and~\citet{Aloisio:2013tda}. According to their findings, the spectral break stands for the transition from a regime where the scattering centers are self-generated by waves generated due to streaming instability to a regime where the instability is damped and particles diffuse in the external Kolmogorov turbulence that cascades from larger spatial scales. 

Indeed, the slope change could be due to the difference in the diffusion processes. But the break does not arise from ion-neutral damping or nonlinear Landau damping as they discussed, but rather from the suppression of streaming instability by background turbulence as first predicted in~\citet{Yan:2002} and confirmed by later studies including~\citet{Farmer:2003mz,Yan:2004,Beresnyak:2008ng}.

\section{Conclusions}
Diffusion models based on a simplified treatment of CR interactions with the environment are not adequate to make predictions in the light of the available/upcoming data. 
From the theoretical point of view, it has been clear that CR propagation in realistic astrophysical turbulence should be revisited in accordance with the recent advances in the understanding of MHD turbulence.  
Based on the test models of turbulence, a series of works carried out by~\citet{Yan:2002,Yan:2004,Yan:2008} have demonstrated that CR diffusion is different from earlier pictures, resulting in a paradigm shift of CR transport theory.
In this work we present for the first time a scenario in which galactic CR propagation is modeled according to the NLT developed by~\citet{Yan:2008} on the basis of a tested model of turbulence.

We can summarize our results as following:
\begin{enumerate}
\item We show that the diffusion coefficient can exhibit the scaling with energy compatible with observations if we adopt tested models of turbulence in which fast modes dominate the scattering of CRs. Accordingly, the dependence of the spatial diffusion coefficient with energy naturally occurs due to the dependence of damping on the local environments.

\item We show that the peak at $\sim 1$~GeV~n$^{-1}$ of the B/C ratio can be reproduced without adding re-acceleration. We aim to further test this scenario by comparing our predictions with the interstellar spectra inferred by accurate diffuse synchrotron and gamma molecular cloud observations.

\item The change of diffusion properties at high energies can be related to the different behavior of diffusion in the galactic plane with respect to the halo determined by local ISM properties. 
Note that a similar explanation has been provided by~\citet{Tomassetti:2012ga} even if based on a purely phenomenological approach.
\end{enumerate}

The results presented in this work can be easily extended to study nonlocal observables, e.g., diffuse gamma and synchrotron emission, in a global model in which diffusion properties are different in the different galactic environments. Our model also allows to distinguish between perpendicular and parallel diffusion, and it will be of extreme interest to test the impact of anisotropic diffusion in a three-dimensional galactic propagation framework, as the one put to work in~\cite{Gaggero2013}.

Understanding the origin of CRs means being able  to unfold a complex chain of physical processes that go from the acceleration of charged particles in still uncertain sources to their propagation in the ISM. 
Each of these steps should be based on a correct understanding of the plasma processes which have been tested.

\section*{acknowledgments}

We thank G. Brunetti, P. Mertsch, R. Schlickeiser and the DRAGON team for fruitful discussions.
CE acknowledges support from the Helmholtz Alliance for Astroparticle Physics funded by the Initiative and Networking Fund of the Helmholtz Association. HY acknowledges the support by NSFC grant AST-11073004 and the visiting fellowship at U Montpellier and Observatoire Midi-Pyren\'ees in Toulouse.

\bibliographystyle{apj}
\bibliography{nltd}
\end{document}